\documentclass[aps,amsfonts,amssymb,preprint,eqsecnum,nofootinbib,superscriptaddress]{revtex4}
\usepackage{epsfig}
\usepackage{amsmath,amsfonts,bm}
\begin{document}

\count255=\time\divide\count255 by 60 \xdef\hourmin{\number\count255}
  \multiply\count255 by-60\advance\count255 by\time
 \xdef\hourmin{\hourmin:\ifnum\count255<10 0\fi\the\count255}

\newcommand{\Dslash}{D\hspace{-0.7em}{ }\slash\hspace{0.2em}}
\newcommand\<{\langle}
\renewcommand\>{\rangle}
\renewcommand\d{\partial}
\newcommand\LambdaQCD{\Lambda_{\textrm{QCD}}}
\newcommand\tr{\mathop{\mathrm{Tr}}}
\newcommand\+{\dagger}
\newcommand\g{g_5}

\newcommand{\xbf}[1]{\mbox{\boldmath $ #1 $}}

\title{Higher-Derivative Lee-Wick Unification}

\author{Christopher D. Carone}
\email{cdcaro@wm.edu}

\affiliation{Department of Physics, College of William \& Mary,
Williamsburg, VA 23187-8795}

\date{April 2009}

\begin{abstract}
We consider gauge coupling unification in Lee-Wick extensions of the Standard Model 
that include higher-derivative quadratic terms beyond the minimally required set.
We determine how the beta functions are modified when some Standard Model particles have 
two Lee-Wick partners.  We show that gauge coupling unification can be 
achieved in such models without requiring the introduction of additional fields 
in the higher-derivative theory and we comment on possible ultraviolet completions.
\end{abstract}


\maketitle

\section{Introduction} \label{intro}
The Lee-Wick Standard Model (LWSM) has been proposed as a possible solution to the hierarchy
problem~\cite{gow}, motivated by the ideas of Ref.~\cite{oldLW}.  For each Standard Model particle,
higher-derivative quadratic terms are introduced so that propagators fall off more quickly with 
momentum.  Although gauge invariance implies that higher-derivative interaction terms must also 
be present, power-counting arguments indicate that the ultraviolet divergences in loop diagrams 
are no greater than logarithmic, even when the usually problematic Higgs sector is taken into 
account~\cite{gow}.

The presence of higher-derivative quadratic terms leads to additional poles in the two-point
function of each Standard Model field.  The higher-derivative theory can be recast using an
auxiliary field approach as a dimension-four Lagrangian, with additional fields corresponding
to the new Lee-Wick partner states~\cite{gow}.  In the original LWSM proposal, each Standard Model 
particle has a single LW partner which, in the dimension-four form of the Lagrangian, has wrong-sign 
kinetic and mass terms.  Due in part to this sign difference, the LW partners play the role of
Pauli-Villars regulators in loop diagrams, so that the cancellation of quadratic divergences
found in the equivalent higher-derivative theory is reproduced.  Unlike Pauli-Villars regulators,
however, Lee-Wick particles are taken to be physical.  It has been argued that Lee-Wick field
theories preserve macroscopic causality as long as the LW partners can decay~\cite{causal}, and that 
gauge boson scattering remains unitary despite the presence of massive LW vector meson 
states~\cite{unitary}.  The evidence in favor of the consistency of LW theories~\cite{causal,unitary,cons} 
and the simple mechanism that they provide for solving the hierarchy problem has motivated a number 
of recent studies of the formal properties and phenomenology of the LWSM and related theories~\cite{big}.

It has been pointed out that LW theories with more than a single LW partner field can be
constructed if higher-derivative quadratic terms beyond the minimally required set are 
included~\cite{CL2}.  Letting $N$ refer to the number of poles in the two-point function of each 
field in the higher-derivative form of the theory, the LWSM most frequently discussed in the 
literature corresponds to $N\!=\!2$; Ref.~\cite{CL2} showed how one may construct the $N\!=\!3$ 
generalization of the LWSM, and provided the mappings between the Lagrangian in its higher-derivative (HD), 
auxiliary field (AF) and Lee-Wick (LW) forms, where the latter refers to the theory with quadratic 
terms that are canonical aside from their overall signs.  Clearly, generalization to LW theories 
with $N\!>\!3$ is possible.  However, one might ask whether anything useful is gained in constructing 
such theories, aside from intellectual exercise.  In Ref.~\cite{CL2}, it was pointed out that the 
heavier LW partner of each Standard Model field in the $N\!=\!3$ theory is an ordinary particle 
(corresponding to a state with positive norm), and therefore might be distinguishable at colliders 
from the lightest LW partner.  In this letter, we point out another, potentially useful feature of 
theories in which some ordinary particles have more than a single LW partner: gauge coupling unification 
can be achieved at the one-loop level without requiring the introduction of new particles that remain 
light in the limit that the LW scale is taken to infinity.

This letter is organized as follows.  In the next section we show how the computation of beta 
functions in the $N\!=\!2$ theory, as considered by Grinstein and O'Connell~\cite{go}, is modified 
in the $N\!=\!3$ case.  Notably, the doubling of the number of the massive LW gauge bosons does not 
lead to a doubling of their contribution to the beta functions, so that one cannot naively 
extrapolate the answer from that of the $N\!=\!2$ theory.  In Section~3 we study one-loop 
unification assuming that each Standard Model particle has either one or two LW partners.  In 
Section~4 we suggest possible extra-dimensional ultraviolet completions for some models of 
this type and we summarize our conclusions.

\section{Beta functions} \label{sec:two}
We employ the background field method, where gauge fields are expanded about a classical
background $B^\mu$,
\begin{equation}
A^\mu \rightarrow B^\mu + A^\mu  \,\,\,,
\end{equation}
where we use the notation $A_\mu \equiv A_\mu^a\, T^a$, etc., with the gauge group generators 
normalized $\mbox{Tr }T^a T^b = \frac{1}{2} \delta^{ab}$. The gauge fixing term is given by
\begin{equation}
{\cal L}_{gf} = -\frac{1}{2 g^2} \mbox{Tr }(D^\mu A_\mu)^2  \,\,\,,
\end{equation}
where the covariant derivative is with respect to the background field
\begin{equation}
D^\mu=\partial^\mu-i B^\mu \,\,\, .
\end{equation}
The gauge-fixed Lagrangian is invariant under a residual gauge symmetry in which $B^\mu$ 
transforms as a gauge field and $A^\mu$ as a matter field in the adjoint representation.
Working to quadratic order in the fluctuating field $A_\mu$ and performing its functional 
integral, one obtains the one-loop effective action for the background
field, including the kinetic term
\begin{equation}
-\frac{1}{2} c_B \mbox{Tr } B_{\mu\nu} B^{\mu\nu} \,.
\end{equation}
The beta function for the gauge coupling can be extracted from the coefficient $c_B$.  This
construction is well-known and discussed in textbooks; we refer the reader to Ref.~\cite{peskin}
for a detailed review, and Ref.~\cite{go} for a discussion of subtleties that can arise in LW
theories.

Grinstein and O'Connell demonstrated in the $N\!=\!2$ LWSM that the same beta functions are
obtained whether one works in the HD or the LW form of the theory~\cite{go}.  We expect the same
to hold true for theories with $N\!>\!2$, though in these cases the HD form of the theory is
more cumbersome for Feynman diagram calculations.  As a consistency check, we will do one example 
in an $N\!=\!3$ theory where it is tractable to compute beta functions in both the HD
and LW forms of the theory: we consider the contribution to the SU(N) beta function from a complex
scalar in the fundamental representation.   For the remaining beta function calculations that 
we need, we work with the simpler LW form of the Lagrangian.

\begin{figure}
\includegraphics[scale=.5]{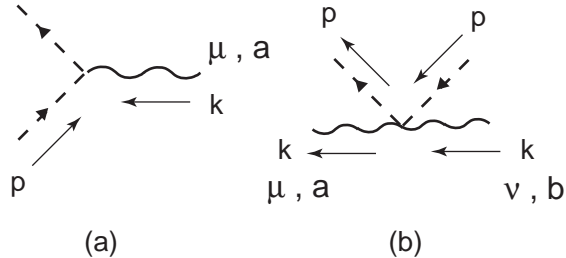}
\caption{\label{fig:verts} Higgs-gauge boson vertices relevant to the calculation
of the gauge boson two-point function, with momenta restricted accordingly.}
\end{figure}
The $N\!=\!3$ Lagrangian for a complex scalar in the fundamental representation of SU(N) is given
by~\cite{CL2}
\begin{equation}
{\cal L}_{\rm HD} = \hat D_\mu \hat{H}^\dagger \hat D^\mu \hat{H}
-m_H^2 \hat{H}^\dagger \hat{H}- \frac{1}{M_1^2}
\hat{H}^\dagger (\hat D_\mu \hat D^\mu)^2 \hat{H} - \frac{1}{M_2^4}
\hat{H}^\dagger (\hat D_\mu \hat D^\mu)^3 \hat{H}  +
{\cal L}_{{\rm int}}(\hat{H})\,\,\,,
\label{eq:higgsl}
\end{equation}
where $\hat{D}_\mu = \partial_\mu -i\, A_\mu-i\, B_\mu$, and the $M_i$ determine the masses of
the LW partners. (We assume that the $M_i$ are comparable and not far above the weak scale.) The 
logarithmically divergent part of $c_B$ determines the beta function.  Equivalently, one can
find the beta function by computing the wave-function renormalization $Z$ of the fluctuating field 
$A$ in background field gauge.  Rescaling the fields so that the gauge coupling appears in the 
covariant derivatives and writing $Z=1+a/\epsilon+\cdots$ in dimensional regularization with
$\epsilon=4-d$, the $\beta$ function is given by~\cite{go}
\begin{equation}
\beta = -\frac{1}{4} g^2 \frac{\partial a}{\partial g} \equiv \frac{b\, g^3}{16 \pi^2}\, .  
\label{eq:betadefs}
\end{equation}
In the present example, the necessary vertices can be extracted from Eq.~(\ref{eq:higgsl}) and are 
shown in Fig.~\ref{fig:verts}.  The three-point coupling shown in Fig.~\ref{fig:verts}a has the 
Feynman rule
\begin{equation}
i \Gamma^{(3)\,a}_\mu(p,k) \equiv i g (2 p + k)_\mu \, T^a \, \left\{1-\frac{1}{M_1^2}[p^2+(p+k)^2]+\frac{1}{M_2^4}[
p^4 + p^2\, (p+k)^2 + (p+k)^4 ]\right\} \,,
\label{eq:tpv}
\end{equation}
with the momenta and indices shown in the diagram.  The four-point coupling shown in
Fig.~\ref{fig:verts}b has the Feynman rule
\begin{eqnarray}
&& i \Gamma^{(4)\,ab}_{\mu\nu}(p,k) \equiv i g^2 \, T^a T^b \left\{ g_{\mu\nu} +\frac{1}{M_1^2}  \left[-2 p^2 g_{\mu\nu} - (2 p +k)_\mu
(2 p +k)_\nu\right] \right.  \nonumber \\
&& \left. + \frac{1}{M_2^4} \left[3 p^4 g_{\mu\nu}+
(2 p^2 + (p+k)^2)(2p+k)_\mu (2p+k)_\nu\right] \right\} +  \Big(
a \leftrightarrow b, k \rightarrow -k  \Big) \, ,
\label{eq:fpv}
\end{eqnarray}
in the simplified case where the momenta are chosen as shown in the diagram (the more general
result will not be required).  Finally, the $\hat{H}$ propagator is given by
\begin{equation}
\tilde{D}(p) = \frac{i}{p^2-m_H^2-p^4/M_1^2+p^6/M_2^4}\, .
\label{eq:prop}
\end{equation}
The one-loop contributions to the gauge boson two-point function are given by
\begin{equation}
i \Pi_{1 \,\mu\nu}^{ab}(k) =
\int \frac{d^4 p}{(2 \pi)^4} \mbox{Tr }[i \Gamma^{(4)\,ab}_{\mu\nu}(p,k) \tilde{D}(p)] \,,
\label{eq:pione}
\end{equation}
for the diagram obtained from Fig.~\ref{fig:verts}b by closing the scalar line, and
\begin{equation}
i \Pi_{2 \,\mu\nu}^{ab} = \int \frac{d^4 p}{(2 \pi)^4} \mbox{Tr }[i \Gamma^{(3)\,a}_\mu(p,k) \tilde{D}(p)
\, i \Gamma^{(3)\,b}_\nu(p+k,-k) \tilde{D}(p+k)] \,,
\label{eq:pitwo}
\end{equation}
for the diagram that is second order in the Fig~\ref{fig:verts}a vertex.  Using the expressions given
in Eqs.~(\ref{eq:tpv}), (\ref{eq:fpv}) and (\ref{eq:prop}),  the logarithmically divergent parts of
$\Pi_{1 \,\mu\nu}^{ab}$ and $\Pi_{2 \,\mu\nu}^{ab}$ may be extracted by expanding the integrands in
powers of $p^{-1}$.  One finds the $\epsilon$ poles
\begin{eqnarray}
i \Pi_{1 \,\mu\nu}^{ab}(k) &\rightarrow& - \frac{i g^2}{16 \pi^2} \left(\frac{2}{\epsilon}\right)
\left[3 \frac{M_2^4}{M_1^2} g_{\mu\nu} + 5 k_\mu k_\nu + k^2 g_{\mu\nu} \right] \delta^{ab} \nonumber  \\
i \Pi_{2 \,\mu\nu}^{ab}(k) &\rightarrow&  \frac{i g^2}{16 \pi^2} \left(\frac{2}{\epsilon}\right)
\left[3 \frac{M_2^4}{M_1^2} g_{\mu\nu} + \frac{11}{2} k_\mu k_\nu + \frac{1}{2} k^2 g_{\mu\nu} 
\right] \delta^{ab} \, .
\end{eqnarray}
which combine to give the desired transverse form
\begin{equation}
i \Pi_{\mu\nu}^{ab} \rightarrow -\frac{1}{2} \frac{i g^2}{16 \pi^2} \left(\frac{2}{\epsilon}\right)
(k^2 g_{\mu\nu}-k_\mu k_\nu) \, \delta^{ab} \,\, .
\end{equation}
From Eq.~(\ref{eq:betadefs}), it then follows that the the contribution to the SU(N) beta function is 
given by $\Delta b = 1/2$. By comparison, the contribution to the SU(2) beta function due to the 
Higgs doublet in the Standard Model is $\Delta b = 1/6$.  In the LW form of the $N\!=\!3$ theory, this 
result is enhanced by a factor of three due to the contribution of the LW partners. (One can check that 
the LW sign changes in vertices and propagators occur in each diagram an even number of times.)  Thus, 
the result for $\Delta b$ computed in the HD form of the $N\!=\!3$ theory reproduces the result of the 
LW form, as one would expect.

One might draw the incorrect conclusion from this example that the contribution to the Standard Model
beta functions from the bosonic LW states is simply enhanced by a factor of $3/2$ in going from the 
$N\!=\!2$ to the $N\!=\!3$ theory.  While this is true for the LW Higgs fields (which couple to the
gauge fields like their Standard Model counterpart, up to signs), it is not true in the LW gauge sector.
The gauge-boson self-interactions in the LW form of an $N\!=\!3$ theory were found in Ref.~\cite{CL2}; 
the couplings of the two LW partners, $A_2$ and $A_3$, to the massless gauge field, $A_1$, in
an SU(N) gauge theory are given by~\cite{CL2}
\begin{eqnarray}
{\cal L}&=&\frac{1}{2} \mbox{Tr}(D_\mu A_{2\nu}-D_\nu A_{2\mu})^2
-\frac{1}{2} \mbox{Tr}(D_\mu A_{3\nu}-D_\nu A_{3\mu})^2 \nonumber \\
&-&\frac{i g}{(m_3^2-m_2^2)} \mbox{Tr }(F_{1\mu\nu}\,[m_3 A_2^\mu -m_2 A_3^\mu , m_3 A_2^\nu -m_2 A_3^\nu])
\, ,\label{eq:gauge1}
\end{eqnarray}
where $m_2$ and $m_3$ are the mass eigenvalues of the LW partners and the covariant derivative here
is given by $D^\mu A_j^\nu = \partial^\mu A_j^\nu - i g [A^\mu_1, A_j^\nu]$ for $j=2,3$.  We can 
write these interactions in a form that more easily allows us to compare the result in the $N\!=\!2$ and 
$N\!=\!3$ theories.  Let ${\bf A}\equiv [A_2, A_3]^T$ and define
 \begin{equation}
\eta=\left(\begin{array}{cc} 1 & 0 \\ 0 & -1 \end{array}\right), \,\,\,\,\,
M^2= \left(\begin{array}{cc} m_2^2 & 0 \\ 0 & m_3^2 \end{array} \right), \,\,\,\,\,
C = \frac{1}{m_3^2-m_2^2} \left(\begin{array}{cc} m_3^2 & -m_2 m_3 \\ -m_2 m_3 & m_2^2
\end{array} \right)  \,\,.
\label{eq:udefs}
\end{equation}
Equation~(\ref{eq:gauge1}) is contained in
\begin{equation}
{\cal L} = \frac{1}{2} \mbox{Tr}\,[(D_\mu {\bf A}_\nu - D_\nu {\bf A}_\mu)^T \eta 
(D^\mu {\bf A}^\nu-D^\nu {\bf A}^\mu)]-\mbox{Tr}\,[{\bf A}_\mu^T \eta M^2 {\bf A}^\mu] 
- 2 i \,g \,\mbox{Tr}\,[F_1^{\mu\nu} {\bf A}_\mu^T C {\bf A}_\nu]\,,
\label{eq:gauge2}
\end{equation}
where we have also included the gauge boson mass terms.
\begin{figure}
\includegraphics[scale=.5]{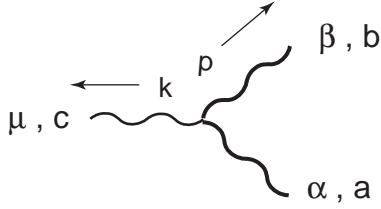}
\caption{\label{fig:gverts} Gauge-LW gauge boson vertex from Eq.~(\ref{eq:gauge2}).  The heavy
line represents the LW gauge boson field ${\bf A}\equiv [A_2, A_3]^T$.}
\end{figure}
The Feynman rule for the $A_1 {\bf A}^2$ vertex shown in Fig.~\ref{fig:gverts}, is
then
\begin{equation}
i \Gamma^{(3)\,abc}_{\alpha\beta\mu}(p,k) = - g f^{abc} \left\{
\eta \left[(2p+k)_\mu g_{\alpha\beta} - p_\alpha g_{\mu\beta}-(p+k)_\beta g_{\mu\alpha}\right]
+C\left[-k_\beta g_{\alpha\mu}+k_\alpha g_{\beta\mu}\right]\right\} \,,
\end{equation}
with the matrices $\eta$ and $C$ as previously defined.  The propagator for the column vector of LW
gauge field partners is given in matrix form by
\begin{equation}
\tilde{D}(p)_{\mu\nu}^{ab} = i \eta [g_{\mu\nu}-M^{-2} p_\mu p_\nu][p^2-M^2]^{-1} \delta^{ab}\,\,\,.
\end{equation}
It is now straightforward to evaluate the $A_1$ two-point function and extract the logarithmically
divergent part, as in our previous scalar example.  We find
\begin{equation}
\Pi_{\mu\nu}^{ab} \rightarrow -i  \frac{\Delta b \, g^2}{16 \pi^2} \left(\frac{2}{\epsilon}\right)
(k^2 g_{\mu\nu}-k_\mu k_\nu) \, \delta^{ab}\,\,,
\end{equation}
where
\begin{equation}
\Delta b=-\frac{1}{2}
\mbox{Tr}\,\left[\openone+6\, C \eta + C \eta C \eta - C \eta M^{-2} C \eta M^2\right] C_2\,\,
\label{eq:ceq}
\end{equation}
and where $C_2$ is the quadratic casimir for the adjoint representation, 
$f^{acd} f^{bcd} = C_2 \, \delta^{ab}$.  Substituting the matrices from Eq.~(\ref{eq:udefs}), one 
obtains $\Delta b=-9/2 \, C_2$.   As a check, we note that in the $N\!=\!2$ theory, where there is a 
single LW partner with mass $m_2$, the appropriate Lagrangian is obtained via the substitutions 
$\openone \rightarrow 1$, $C=\eta=1$ and $M=m_2^2$. In this case, Eq.~(\ref{eq:ceq}) yields 
$\Delta b=-7/2 \, C_2$, in agreement with the result quoted in Ref.~\cite{go}.  The remaining pure 
gauge contribution from the light field $A_1$ and ghosts yields $\Delta b = -11/3 \, C_2$ as in the 
$N\!=\! 2$ theory.  We note that the contribution of the massive LW states in the $N\!=\!3$ theory is 
not twice the $N\!=\!2$ result due to the third term of Eq.~(\ref{eq:gauge1}), which leads to a loop 
diagram in which both the fields $A_2$ and $A_3$ propagate.

\section{One-loop gauge unification}
We may now apply the results of the previous section to evaluate one-loop gauge coupling unification.
For the pure gauge contributions to the beta functions $b_i$ of the gauge group factor $G_i$, the 
Standard Model results are $-11/3 \, C_2 (G_i)$; these are modified to either $-43/6 \, C_2(G_i)$ or 
$-49/6 \, C_2(G_i)$ in the $N\!=\!2$ and $N\!=\!3$ LW extensions, respectively, following the 
discussion in the previous section. The contribution to the $b_i$ for each chiral matter field 
is multiplied by either $3$ or $5$, since there are one or two Dirac partners in the $N\!=\!2$ 
and $N\!=\!3$ LW extension, respectively.  Finally, the Higgs field contribution is multiplied by either 
$2$ or $3$, since each LW partner is also complex scalar.  Ref.~\cite{go} notes that the LWSM 
does not unify at one loop, unless multiple Higgs doublets are included.  As Table~\ref{table:gu}
indicates, we find this is the case if $8$ Higgs doublets are included in the $N\!=\!2$ theory, or $6$ in
$N\!=\!3$.  However, we can now consider models in which each field has at least one LW partner, with 
some having two.  These models solve the hierarchy problem since they are at least as convergent as the 
$N\!=\!2$ theory. This provides a wide range of possibilities for achieving more accurate unification.  In 
Table~\ref{table:gu} we give some of the simpler successful models, with the SM and MSSM one-loop
results provided for comparison.  The experimental central values of  $\alpha_1^{-1}(m_Z)=59.00$ 
and $\alpha_2^{-1}(m_Z)=29.57$~\cite{PDG} are taken as inputs, unification is assumed and 
$\alpha_3^{-1}(m_Z)$ is then predicted.
\begin{table}
\begin{center}
\caption{Predictions for $\alpha_3^{-1}(m_Z)$ assuming one-loop unification. The experimental
value is $8.2169 \pm 0.1148$~\cite{PDG}. The GUT scale is defined by 
$\alpha_1^{-1}(M_{GUT})=\alpha_2^{-1}(M_{GUT})$. The abbreviations used are as follows: H=Higgs doublets,
gen.=generation, LH=left handed.}
\label{table:gu}
\vspace{1em}
\begin{tabular}{cc cc   cccccc}\hline\hline
model  & \quad & $N\!=\!3$ fields  & \quad&    $(b_3,b_2,b_1)$ & $M_{GUT}$ (GeV) 
& \quad & $\alpha_3^{-1}(m_Z)$ &\quad & error\\  \hline
SM     &       & - &&$(-7, -19/6, 41/10)$& $1\times 10^{13}$&  & $14.04$ & & $+50.8 \sigma$ \\
MSSM   &       & - &&$(-3,1,33/5)$        &$2\times 10^{16}$&  & $ 8.55$ & & $+2.9 \sigma$ \\
$N\!=\!2$ 1H LWSM & &none &  & $(-19/2,-2,61/5)$&  $4\times 10^{7}$  &  & $14.03$ & & $+50.6 \sigma$ \\
$N\!=\!3$ 1H LWSM & &all  &  & $(-9/2, 25/6,203/10)$ &$9\times 10^{6}$& & $13.76$ & & $+48.3 \sigma$ \\
$N\!=\!2$ 8H LWSM & &none & & $(-19/2,1/3,68/5)$ &$1\times 10^{8}$& & $7.76$  & & $-4.01 \sigma$ \\
$N\!=\!3$ 6H LWSM &   &   all &&$(-9/2, 20/3, 109/5)$ &$2\times 10^{7}$& & $7.85$ & & $-3.16 \sigma$ \\
$N\!=\!2$ 1H LWSM, & &gluons & &$(-25/2,-2,61/5)$ &$4\times 10^{7}$& & $7.81$ & &  $-3.55\sigma$ \\
$N\!=\!2$ 1H LWSM && gluons, 1 gen. quarks & & $(-59/6,0,41/3)$ &$7\times 10^{7}$& & $8.40$ & & $+1.55\sigma$\\
$N\!=\!2$ 1H LWSM && 1 gen. LH fields & & $(-49/6,2/3,191/15)$ &$4\times 10^{8}$& & $8.03$  & & $-1.66\sigma$ \\
$N\!=\!2$ 2H LWSM &&  LH leptons & & $(-19/2,1/3,68/5)$ &$1\times 10^{8}$& & $7.76$ & & $-4.01 \sigma$ \\
$N\!=\!2$ 2H LWSM && gluons, quarks, 1H & & $(-9/2,9/2,169/10)$ &$3\times 10^{8}$& & $8.21$ & & $-0.06\sigma$\\
\hline\hline
\end{tabular}
\end{center}
\end{table}
Of course, Table~\ref{table:gu} does not represent an exhaustive list of the possible variations on
the LWSM.   It illustrates that models with improved gauge coupling unification at the one-loop level
can be achieved in the higher-derivative LW theories of Ref.~{\cite{CL2}} by choosing an appropriate
set of higher-derivative terms, beyond the minimally required set, without adding additional fields 
in the HD theory.  It should be noted that the results in Table~\ref{table:gu} will be altered by 
two-loop corrections to the running of the gauge couplings, which have not been computed in any 
version of the LWSM.  In addition, specific models will have threshold corrections that will modify
these results.  It should be understood that the deviations from the experimental value of 
$\alpha^{-1}_3(m_Z)$ shown in the table are subject to these uncertainties.

\section{Completions and Conclusions}

Although we will not attempt to construct explicit ultraviolet completions that are consistent
with the LW theories listed in Table~\ref{table:gu}, a number of points are worth noting.  First, the 
unification shown assumes the GUT normalization of hypercharge, the choice that leads to unification 
in conventional SU(5) or trinified gauge theories.  Nevertheless, it is possible in strongly coupled
string theories for the string and unification scales to coincide, so that one may never realize a
grand unified field theory at any intermediate point.  If one were interested in conventional
grand unification, then two issues become relevant.  First, the LW theories in Table~\ref{table:gu}
unify at a scale much lower than in the Standard Model, with the GUT scale ranging from $4 \times 10^7$
to $4 \times 10^8$~GeV in the more successful models.  Ref.~\cite{go} points out that the low unification 
scale in their multi-Higgs LWSM is not consistent with semi-simple unification, due to the constraint 
from proton decay.  However, this assessment may be overly pessimistic.  Higher-dimensional SU(5) GUTs can 
avoid the problem of proton decay from GUT gauge boson exchange by placing fermions at orbifold fixed points 
where the wave functions of the offending bosons vanish (see, for example, the discussion in~\cite{ddg}).  
There does not seem to be any reason why the same approach couldn't be adapted here.  The compactification scale 
in theories where GUT symmetry is broken by orbifold projection can be taken at or near the grand 
unification scale (as in Ref.~\cite{hm}), so the effective theory at lower energies is four-dimensional; the 
beta functions shown in Table~\ref{table:gu} therefore apply, as does the accounting of divergences in 
four-dimensional LW theories.  (Unification in theories with a lower compactification scale would require 
a different analysis since the gauge coupling running is affected significantly by Kaluza-Klein thresholds.)  
The advantage of higher-dimensional GUTs is that one can place incomplete multiplets of matter fields at 
orbifold fixed points where the GUT symmetry is broken.  At such fixed points, it is consistent with gauge 
invariance to write down different higher-derivative kinetic terms for what would otherwise be different 
components of a single GUT multiplet in a 4D theory.  This approach makes it feasible, for example, to have  
an $N\!=\!2$ LW unified theory where only the left-handed fermions of one generation have $N\!=\!3$ 
partners (i.e., the third from last example in Table~\ref{table:gu}).  Finally, one may pursue trinification, 
as advocated in Ref.~\cite{go}, so that there is no gauge-boson-induced proton decay.  In this case, the 
extra-dimensional construction has similar benefits.  In the $N\!=\!3$ six-Higgs doublet model, for example, 
one does not need to introduce six complete {\bf 27}-plets if the GUT group is broken by extra-dimensional 
boundary conditions on an interval, an approach discussed in Refs.~\cite{cartrin}.  It is also worth noting 
that in trinified theories where the equality of SU(3) gauge couplings at the unification scale is a 
consequence of string boundary conditions rather than a discrete cyclic symmetry of the field 
theory~\cite{willen}, the presence of $N\!=\!3$ gluons would be consistent with the SU(3)$^3$ gauge 
symmetry and would allow unification without a large multiplicity of Higgs doublets.  

In summary, we have shown that the particle content needed to fix one-loop gauge unification in the LWSM 
can be introduced in a more restricted way than previously considered, by extending the non-generic set 
of HD interactions that are consistent with the LW construction to higher order for some Standard Model
fields.  Computation of the pure gauge contributions to the beta functions requires a computation that
does not seem to generalize trivially to theories with arbitrary $N$, and was computed here for the 
next-to-minimal case of $N\!=\!3$.  Explicit unified field theories that correspond to some of the 
solutions discussed in the previous section seem plausible in the framework of orbifold GUTs, where matter 
fields may be placed at fixed points with reduced gauge symmetry so that HD kinetic terms may differ
between fields that would otherwise live within the same 4D GUT multiplet.  The construction of explicit 
unified theories of this type seems worthy of further investigation.
 
\vspace{-0.5em}
\section*{Acknowledgments}
\vspace{-0.5em}
This work was supported by the NSF under Grant Nos.\ PHY-0456525 and PHY-0757481.  We thank
Rich Lebed for a careful reading of the manuscript and useful comments.


\begin{thebibliography}{99}

\bibitem{gow}
  B.~Grinstein, D.~O'Connell and M.B.~Wise,
  Phys.\ Rev.\  D {\bf 77}, 025012 (2008)
  [arXiv:0704.1845 [hep-ph]].  
\bibitem{oldLW}
  T.~D.~Lee and G.~C.~Wick,
  Nucl.\ Phys.\  B {\bf 9}, 209 (1969);
  Phys.\ Rev.\  D {\bf 2}, 1033 (1970).
\bibitem{causal}
  B.~Grinstein, D.~O'Connell and M.~B.~Wise,
  arXiv:0805.2156 [hep-th].
\bibitem{unitary}
  B.~Grinstein, D.~O'Connell and M.~B.~Wise,
  Phys.\ Rev.\  D {\bf 77}, 065010 (2008)
  [arXiv:0710.5528 [hep-ph]].
\bibitem{cons}
  K.~Jansen, J.~Kuti and C.~Liu,
  Phys.\ Lett.\  B {\bf 309}, 119 (1993)
  [arXiv:hep-lat/9305003];
  Phys.\ Lett.\ B {\bf 309}, 127 (1993)
  [arXiv:hep-lat/9305004];
Z.~Fodor, K.~Holland, J.~Kuti, D.~Nogradi and C.~Schroeder,
  PoS {\bf LAT2007}, 056 (2007)
  [arXiv:0710.3151 [hep-lat]];
  C.~Liu,
  arXiv:0704.3999 [hep-ph];
A.~van Tonder,
  Int.\ J.\ Mod.\ Phys.\  A {\bf 22}, 2563 (2007)
  [arXiv:hep-th/0610185];
  arXiv:0810.1928 [hep-th].
\bibitem{big}
T.~G.~Rizzo,
  JHEP {\bf 0706}, 070 (2007)
  [arXiv:0704.3458 [hep-ph]];
  {\bf 0801}, 042 (2008)
  [ar\-Xiv:0712.1791 [hep-ph]];
C.~D.~Carone and R.~F.~Lebed,
  Phys.\ Lett.\  B {\bf 668}, 221 (2008)
  [arXiv:0806.4555 [hep-ph]];
E.~Alvarez, L.~Da Rold, C.~Schat and A.~Szynkman,
  JHEP {\bf 0804}, 026 (2008)
  [arXiv:0802.1061 [hep-ph]];
  arXiv:0810.3463 [hep-ph];
  J.~R.~Espinosa, B.~Grinstein, D.~O'Connell and M.~B.~Wise,
  Phys.\ Rev.\  D {\bf 77}, 085002 (2008)
  [arXiv:0705.1188 [hep-ph]];
E.~Gabrielli,
  Phys.\ Rev.\  D {\bf 77}, 055020 (2008)
  [arXiv:0712.2208 [hep-ph]];
F.~Wu and M.~Zhong,
  Phys.\ Lett.\  B {\bf 659}, 694 (2008)
  [arXiv:0705.3287 [hep-ph]];
F.~Krauss, T.~E.~J.~Underwood and R.~Zwicky,
  Phys.\ Rev.\  D {\bf 77}, 015012 (2008)
  [arXiv:0709.4054 [hep-ph]];
 T.~E.~J.~Underwood and R.~Zwicky,
  Phys.\ Rev.\  D {\bf 79}, 035016 (2009)
  [arXiv:0805.3296 [hep-ph]];
T.~R.~Dulaney and M.~B.~Wise,
  Phys.\ Lett.\  B {\bf 658}, 230 (2008)
  [arXiv:0708.0567 [hep-ph]];
F.~Knechtli, N.~Irges and M.~Luz,
  arXiv:0711.2931 [hep-ph];
F.~Wu and M.~Zhong,
  Phys.\ Rev.\  D {\bf 78}, 085010 (2008)
  [arXiv:0807.0132 [hep-ph]];
  S.~Lee,
  arXiv:0810.1145 [astro-ph];
  A.~M.~Shalaby,
  arXiv:0812.3419 [hep-th];
  A.~Rodigast and T.~Schuster,
  arXiv:0903.3851 [hep-ph];
  B.~Fornal, B.~Grinstein and M.~B.~Wise,
  arXiv:0902.1585 [hep-th].
\bibitem{CL2}
  C.~D.~Carone and R.~F.~Lebed,
  JHEP {\bf 0901}, 043 (2009)
  [arXiv:0811.4150 [hep-ph]].
\bibitem{go}
 B.~Grinstein and D.~O'Connell,
  Phys.\ Rev.\  D {\bf 78}, 105005 (2008)
  [arXiv:0801.4034 [hep-ph]];
\bibitem{peskin}
  M.~E.~Peskin and D.~V.~Schroeder, ``An Introduction To Quantum Field Theory,''
{\it  Reading, USA: Addison-Wesley (1995) 842 p}.
\bibitem{PDG}
  C.~Amsler {\it et al.}  [Particle Data Group],
  Phys.\ Lett.\  B {\bf 667}, 1 (2008).
\bibitem{ddg}
K.~R.~Dienes, E.~Dudas and T.~Gherghetta,
  Nucl.\ Phys.\  B {\bf 537}, 47 (1999)
  [arXiv:hep-ph/9806292].
\bibitem{hm}
A.~Hebecker and J.~March-Russell,
  Phys.\ Lett.\  B {\bf 541}, 338 (2002)
  [arXiv:hep-ph/0205143].
\bibitem{cartrin}
  C.~D.~Carone and J.~M.~Conroy,
  Phys.\ Rev.\  D {\bf 70}, 075013 (2004)
  [arXiv:hep-ph/0407116].
\bibitem{willen}
  S.~Willenbrock,
  Phys.\ Lett.\  B {\bf 561}, 130 (2003)
  [arXiv:hep-ph/0302168].
\end{thebibliography}
\end{document}